\documentclass[aps,prl,twocolumn,groupaddress]{revtex4-1}

\usepackage{graphicx}
\usepackage[rightcaption]{sidecap}
\usepackage{floatrow}
\usepackage{amsmath}
\usepackage{array}
\usepackage{color}
\usepackage{xfrac}
\newcolumntype{@}{>{\global\let\currentrowstyle\relax}}
\newcolumntype{^}{>{\currentrowstyle}}

\newcommand{\ssymbol}{$6^2 \text{S}_{1/2}$}
\newcommand{\psymbol}{$6^2 \text{P}_{1/2}$}
\newcommand{\dsymbol}{$5^2 \text{D}_{3/2}$}
\newcommand{\otherdsymbol}{$5^2 \text{D}_{5/2}$}
\newcommand{\ba}{$^{133}\text{Ba}^+$}

\begin{document}


\title{Spectroscopy of a synthetic trapped ion qubit}


\author{David Hucul, Justin E. Christensen,  Eric R. Hudson,  Wesley C. Campbell}
\affiliation{Department of Physics and Astronomy, University of California -- Los Angeles, Los Angeles, California, 90095, USA}

\date{\today}

\begin{abstract}
\ba\ has been identified as an attractive ion for quantum information processing due to the unique combination of its spin-1/2 nucleus and visible wavelength electronic transitions. Using a microgram source of radioactive material, we trap and laser-cool the synthetic \textit{A} = 133 radioisotope of barium II in a radio-frequency ion trap. Using the same, single trapped atom, we measure the isotope shifts and hyperfine structure of the \psymbol\ $\leftrightarrow$ \ssymbol\ and \psymbol\ $\leftrightarrow$ \dsymbol\ electronic transitions that are needed for laser cooling, state preparation, and state detection of the clock-state hyperfine and optical qubits. We also report the \psymbol\ $\leftrightarrow$ \dsymbol\ electronic transition isotope shift for the rare \textit{A} = 130 and 132 barium nuclides, completing the spectroscopic characterization necessary for laser cooling all long-lived barium II isotopes.
\end{abstract}


\maketitle

Since the demonstration of the first CNOT gate over 20 years ago \cite{Monroe1995}, trapped ion quantum information processing (QIP), including  quantum simulation,  has developed considerably  \cite{blatt:2008}, recently demonstrating fully-programmable quantum processors \cite{hanneke:2010, debnath:2016}. To date, qubits have been demonstrated in trapped ion hosts of all non-radioactive, alkaline-earth-like elements \cite{Monroe1995, tan:2015, nagerl:2000, keselman:2011, dietrich:2008, dietrich:2010, lee:2003, balzer:2006, olmschenk:2007}. These ions possess a simple electronic structure that facilitates straightforward laser cooling as well as quantum state preparation, manipulation, and readout via electromagnetic fields.

For the coherent manipulation of qubits, the phase of this applied electromagnetic field must remain stable with respect to the qubit phase evolution. Thus, atomic hyperfine structure is a natural choice for the definition of a qubit, as these extremely long-lived states can be manipulated with easily-generated, phase-coherent microwave radiation. In particular, qubits defined on the hyperfine structure of ions with half-integer nuclear spin possess a pair of states with no projection of the total angular momentum ($F$) along the magnetic field ($m_F = 0$). These so-called ``clock-state" qubits are well-protected from magnetic field noise and can yield coherence times exceeding 10 minutes \cite{fisk:1997, wang:2017}. Further, for these species, $F = 0$ ground and excited states only occur when the nuclear spin $I=1/2$. This is desirable because the $F^{\prime}=0 \not\leftrightarrow F^{\prime \prime}= 0$ selection rule can be leveraged to produce fast, robust qubit state preparation and readout that relies solely on frequency selectivity \cite{lee:2003, olmschenk:2007}.

Among the alkaline-earth-like elements, only three (Cd, Hg, Yb) have naturally occurring $I = 1/2$ isotopes. Mercury and cadmium ions require lasers in the deep ultraviolet portion of the electromagnetic spectrum, making it difficult to integrate them into a large-scale QIP architecture. Since $^{171}\text{Yb}^+$ has the longest laser-cooling wavelength at 370 nm, it has been used in a wide variety of groundbreaking QIP experiments  \cite{islam:2013, richerme:2014, debnath:2016, hucul:2015, zhang:2016, neyenhuis:2016}. However, even at this ultraviolet wavelength, the use of photonics infrastructure developed for visible and infrared light is limited. For example, significant fiber attenuation limits the long-distance transmission of quantum information at 370 nm. Furthermore, in $\text{Yb}^+$, the short lifetime of the ${5}^2\text{D}_{5/2}$ manifold (7 ms  \cite{taylor:1997}), along with decays to a low-lying ${}^2\text{F}_{7/2}$ manifold, complicate state-selective shelving of the hyperfine qubit with ultra-high fidelity readout and direct manipulation of an optical qubit \cite{huntemann:2012, harty:2014}.

\begin{figure*}[!t]
\includegraphics[width=\columnwidth]{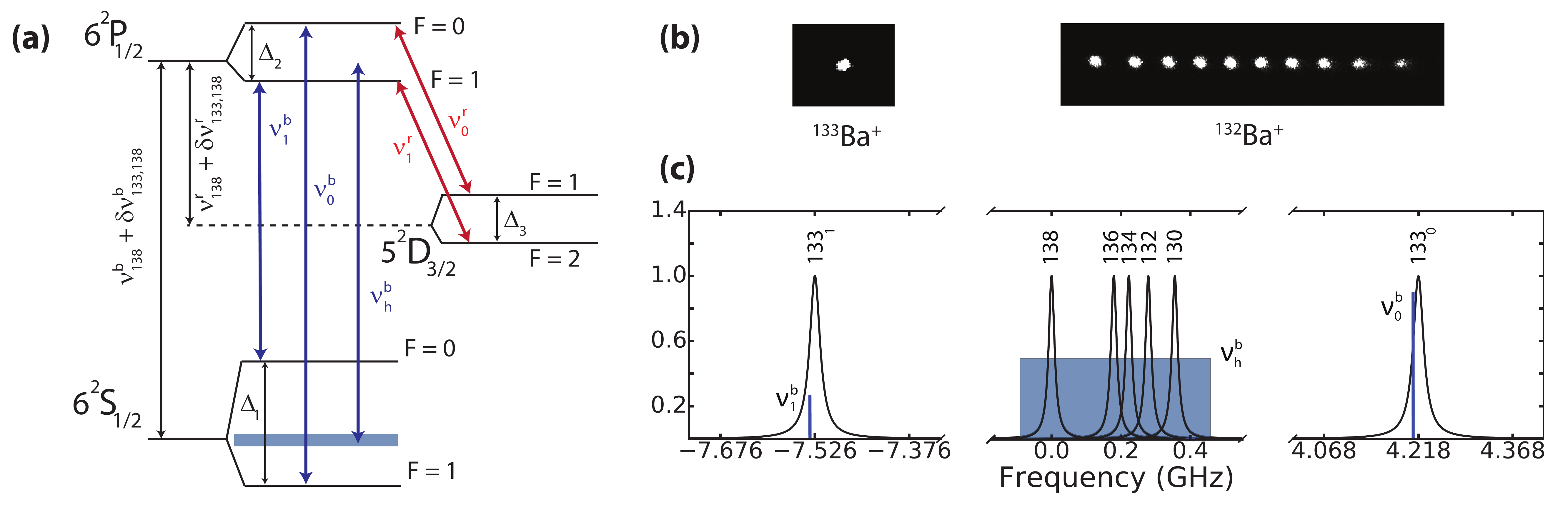}
\caption{\label{fig:1} \textbf{(a)} Laser cooling transitions for the $A = 133$ isotope of barium II with hyperfine structure of the underlying states. \textbf{(b)} A single \ba\ ion and an isotopically pure  $^{132}\text{Ba}^+$ ion chain loaded from an enriched microgram source of barium atoms. \textbf{(c)} Laser loading scheme of \ba\ for the \ssymbol\ $\leftrightarrow$ \psymbol\ transition. To Doppler cool \ba, the laser carrier $\nu_{0}^{b}$ is stabilized 4.218(10) GHz above the $^{138}\text{Ba}^+$ resonance. The frequency $\nu_{1}^{b}$, resulting from a second-order sideband at $\nu_{0}^b - 11.744$~GHz, depopulates the \ssymbol, $F$ = 0 state. The frequency $\nu_{h}^{b}$, resulting from a first-order sideband at $\nu_{0}^b -4.300$~GHz, Doppler cools any co-trapped barium II even isotopes and sympathetically cools \ba. This first-order sideband is scanned across the blue shaded region (to $\nu_{0}^b -3.800$~GHz)  using a high bandwidth fiber EOM to Doppler heat any other barium II isotopes out of the  ion trap.}
\end{figure*}

A possible remedy to these problems exists in the synthetic $A = 133$ isotope of barium ($\tau_{1/2} = 10.5$ years), which combines the advantages of many different ion qubits into a single system. \ba\ has nuclear spin $I=1/2$, allowing fast, robust state preparation and readout of the hyperfine qubit; metastable D states ($\tau$ $\approx$ 1~min), allowing  ultra-high fidelity readout \cite{harty:2014}; and long-wavelength transitions enabling the use of photonic technologies developed for the visible and near infrared spectrum.

Here, we demonstrate loading and laser-cooling of \ba\ atomic ions from a microgram source of barium atoms. We measure the previously unknown \ba\ isotope shift of the \psymbol\ $\leftrightarrow$ \dsymbol\ transition and the  hyperfine constant of the \dsymbol\ state. Our measurements of the other spectroscopic features of \ba\ are in agreement with earlier measurements \cite{harmatz:1966, hohle:1976, knab:1987}. In addition, using the same techniques, we measure and report the isotope shifts of the \psymbol\ $\leftrightarrow$ \dsymbol\ transition in the rare $^{130}\text{Ba}^+$ and $^{132}\text{Ba}^+$ species. 
 
\begin{figure*}[t!]
\includegraphics[width=\columnwidth]{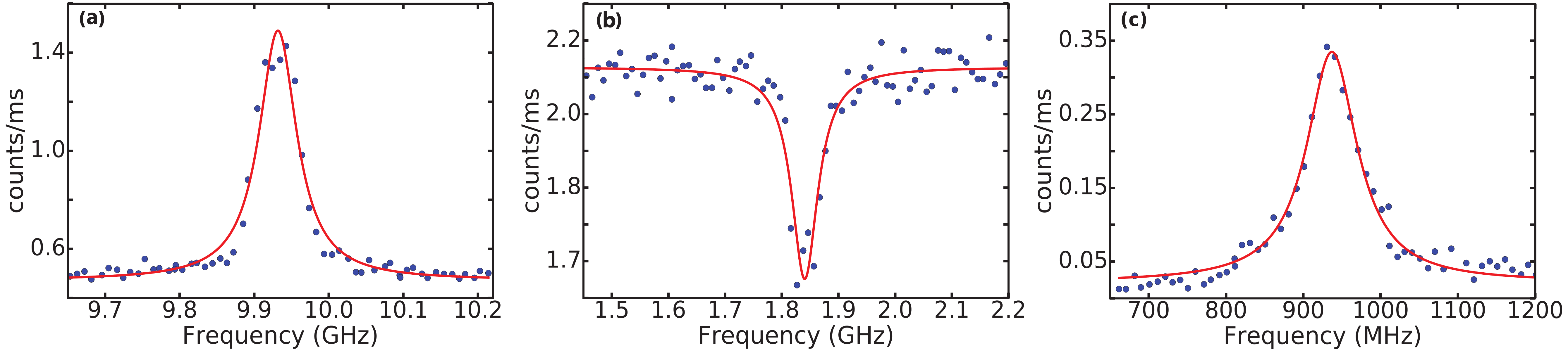}
\caption{\label{fig:2} Measured hyperfine splittings of the \ssymbol, \psymbol, and \dsymbol\  states in $^{133}$Ba$^+$. Solid red lines  are fitted Lorenztian profiles. \textbf{(a)} Fluorescence from a single \ba atomic ion with the application of laser frequencies $\nu_{0}^{b}$ and $\nu_{0,1}^{r}$ while scanning laser frequency $\nu_{1}^{b}$. The peak of the fluorescence spectrum yields the \ssymbol\ hyperfine qubit splitting $\Delta_1 = 9931(2)_\text{stat}$ MHz. \textbf{(b)} Fluorescence as a function of applied modulation frequency to a laser tuned slightly red of the \ssymbol, $F = 1$ $\leftrightarrow$ \psymbol, $F = 0$ transition. When the applied modulation frequency is near $\Delta_2$, the ion can spontaneously decay to the $F = 2$ states in the \dsymbol\ manifold. The resulting decrease in fluorescence gives a \psymbol\ hyperfine splitting of $\Delta_2 = 1840 (2)_\text{stat}$ MHz.  \textbf{(c)} After applying laser frequencies $\nu_{0,1}^b$, $\nu_{0}^{b} - \Delta_2$, and $\nu_{0}^{r}$,  an applied frequency near $\nu_{1}^{r}$  is scanned to repump \ba\ out of the $F = 2$ states in the \dsymbol\ manifold. The resulting increase in fluorescence rate yields a \dsymbol\ hyperfine splitting of $\Delta_3 = 937 (3)_\text{stat}$ MHz. These measurements all have a $\pm$20 MHz systematic uncertainty primarily due to drift of the wavemeter used to stabilize the lasers.}
\end{figure*}

For this work, barium ions are confined using a linear radio frequency (rf) Paul trap. The minimum distance between the trap axis and the electrodes is 3 mm and the trap operates with a peak-to-peak rf voltage $V_{pp}= 200$~V at frequency $\Omega \approx 2\pi\times$1~MHz. Each electrode can be independently DC biased allowing for the compensation of stray fields and the ejection of trapped ions into a laser-cooling-assisted mass spectrometer (LAMS) \cite{wiley:1955, schneider:2014}. Laser cooling of barium ions is accomplished with wavelengths near 493~nm and 650~nm. These lasers enter separate fiber electro-optic modulators (EOMs) with 6~GHz bandwidth and are delivered to the experiment via single-mode optical fibers. The EOMs are used to provide frequency sidebands on the laser spectrum, which allow cooling and/or heating multiple isotopes simultaneously, as well as for addressing the necessary transitions due to hyperfine structure in $I\neq 0$ isotopes (see Table \ref{tab:iso shifts}). An applied magnetic field of a few Gauss along with  laser beams linearly polarized  $\approx$ 45$^{\circ}$ from the magnetic field direction are used to destabilize dark states that result from coherent population trapping \citep{berkeland:2002}.

A source of $^{133}$Ba atoms is produced by drying a commercially available  solution of neutron activated BaCl$_2$ dissolved in 0.1~M HCl on a platinum ribbon substrate. The vendor reports that approximately  2\% of the total barium atoms are $^{133}$Ba \cite{ez:2016}. Atomic ion fluorescence and a LAMS spectrum indicate a highly enriched source of $^{132}\text{Ba}$ atoms due to the manufacturing process. The platinum ribbon substrate is $\approx$ 4~mm  from the edge of the trap in the radial direction, and near the center of the trap axially. A 532~nm, 0.4~mJ, 5-7~ns laser pulse produces ions by ablating the barium on the platinum ribbon substrate. Turning on the rf voltage 10~$\mu$s after laser ablation confines ions in the ion trap.  Overlapped cooling and repumping beams enter the trap at an angle of $45^\circ$ and $0^\circ$ with respect to the axial direction of the ion trap.

\begin{table*}[t]
\caption{Isotope shifts of the \psymbol\ $\leftrightarrow$ \ssymbol\ and \psymbol\ $\leftrightarrow$ \dsymbol\ electronic transitions of barium atomic ions and hyperfine $\mathcal{A}$ and $\mathcal{B}$ constants.  The isotope shift of the $i$-th electronic transition is defined relative to $^{138}\text{Ba}^+$ and is $\delta \nu^{i} \equiv \nu_{A}^i - \nu_{138}^i$. The isotope shifts of all barium atomic ions are positive with the exception of the isotope shift of the \psymbol\ $\leftrightarrow$ \dsymbol\  transition in $^{137}\text{Ba}^+$. The bolded values are spectroscopic measurements from this work and have a systematic uncertainty of $\pm$20 MHz. All other isotope shifts are reported from references \citep{hohle:1976, blatt:1982, wendt:1984, vanhove:1985, knab:1987, villemoes:1993}. Columns 3-8 are in MHz.}\label{tab:iso shifts}
\begin{ruledtabular}
\begin{tabular}{cccccccc}
	$A$	& $I$	& 	$\delta \nu^{b}$	& $\delta \nu^{r}$	& $\mathcal{A}_{\text{S}_{1/2}}$		& $\mathcal{A}_{\text{P}_{1/2}}$	& $\mathcal{A}_{\text{D}_{3/2}}$	& $\mathcal{B}_{\text{D}_{3/2}}$ \\	
	\hline	 
	130	& 0				&	355.3(4.4)	& $\textbf{394(1)}_{\text{stat}}$	& -						& -					&	-									& -			\\
	132	& 0				&	278.9(4)	& $\textbf{292(1)}_{\text{stat}}$	& -						& -					&	-									& -			\\
	133	&\sfrac{1}{2}	&	373(4)		& $\textbf{198(4)}_{\text{stat}}$	& -9925.45355459(10)	& -1840(11)			&	$\textbf{-468.5(1.5)}_{\text{stat}}$	& -			\\
	134	& 0				&	222.6(3)	& 174.5(8)							& -						& -						&	-									& -			\\
	135	&\sfrac{3}{2}	&	348.6(2.1)	& 82.7(6)							& 3591.67011718(24)		& 664.6(3)			&	169.5892(9)							& 28.9536(25)\\
	136	& 0				&	179.4(1.8)	& 68.0(5)							& -						& -					&	-									& -			\\
	137	&\sfrac{3}{2}	&	271.1(1.7)	& -13.0(4)							& 4018.87083385(18)		& 743.7(3)			&	189.7288(6)							& 44.5417(16)\\
	138	& 0				&	$\equiv$ 0 	& $\equiv$ 0						& -						& -					&	-									& -			\\
\end{tabular}
\end{ruledtabular}
\end{table*}

To Doppler cool \ba, a laser near 493 nm is slightly red-detuned ($\approx$ 30 MHz) from the \psymbol, $F$ = 0 $\leftrightarrow$ \ssymbol, $F$ = 1  transition, denoted $\nu_{0}^b$ in Fig. \ref{fig:1}a. Transitions between the \psymbol, $F$ = 0 $\leftrightarrow$ \ssymbol, $F$ = 0 are forbidden, but off-resonant scattering via the \psymbol, $F = 1$ states leads to population trapping in the \ssymbol, $F$ = 0 state. To depopulate this state, the 493~nm fiber EOM is driven at $\nu_0 = 5.872$ GHz resulting in a second-order sideband resonant with  the \psymbol, $F$ = 1 $\leftrightarrow$ \ssymbol, $F$ = 0 transition. A re-pumping laser near 650 nm  is slightly red-detuned of the  \psymbol, $F$ = 0 $\leftrightarrow$ \dsymbol, $F$ = 1 transition, denoted $\nu_{0}^r$ (see Fig. \ref{fig:1}a). Transitions between the \psymbol, $F$ = 0 $\leftrightarrow$ \dsymbol, $F$ = 2 are dipole-forbidden, but decay from the  \psymbol, $F=1$ states populates the \dsymbol, $F$ = 2 states. The off-resonant scatter rate out of the \dsymbol, $F=2$ states, from the applied laser frequency $\nu_{0}^r$,  is greater than the decay rate into the state due to off-resonant scatter from the application of laser frequency $\nu_{0}^b$. Therefore, only the three frequencies $\nu_{0}^b$, $\nu_{1}^b$, and $\nu_{0}^{r}$ are required to cool and crystallize \ba. To improve cooling,  the 650~nm fiber EOM is driven at 904 MHz resulting in a first order sideband red-detuned from the  \psymbol, $F$ = 1 $\leftrightarrow$ \dsymbol, $F = 2$ transition, denoted $\nu_{1}^r$ in Fig. \ref{fig:1}a.

During laser ablation, other ions (here, mainly $^{132}\text{Ba}^+$ due to their high abundance in our source) tend to be co-trapped with \ba. Because the \ssymbol\ hyperfine qubit splitting of \ba\ is much larger than the isotope shift of the \psymbol\ $\leftrightarrow$ \ssymbol\  transition in all $\text{Ba}^+$ isotopes, we are able to utilize a single high bandwidth fiber EOM to simultaneously laser cool \ba\ while laser-heating any even barium isotopes out of the ion trap (see Fig. \ref{fig:1}c). Additional laser sidebands can be used to laser-heat the odd isotopes out of the ion trap using the \psymbol\ $\leftrightarrow$ \dsymbol\ transitions, although in practice infrequent loading rates of these species from the neutron activated BaCl$_2$ microgram source rarely require this. Other chemical species with significantly different charge to mass ratio can be ejected from the ion trap by ramping the trap voltages. $^{133}\text{Ba}$ decays to form $^{133}\text{Cs}$ with a half-life of 10.5 years. Since $^{133}\text{Ba}$ and $^{133}\text{Cs}$ have nearly identical masses, trap filtration based on charge to mass ratio cannot be used to separate them. By monitoring thermionic emission from a heated platinum filament, we find that $^{133}\text{Cs}$ can easily and regularly be preferentially removed from a Ba source in situ. 

The technique of isotopic purification via isotope-selective heating appears to be effective at removing unwanted ions without any observable loss of the desired species. Detailed molecular dynamics simulations of the process have not revealed any  loss of the target ion, even when co-trapped with 499 ions undergoing laser-heating. This is critical for working with radioactive isotopes as it allows the use of non-isotope-selective loading techniques, like laser ablation, to be used. 

As shown in Fig. \ref{fig:1}a, the magnetic moment of the $I=1/2$ \ba\ nucleus splits each fine-structure state by  $\mathcal{H} = \mathcal{A} \vec{I} \cdot \vec{J}$, where $\mathcal{A}$ is the magnetic hyperfine constant associated with each fine structure state. The hyperfine splittings of the \ssymbol, \psymbol, and \dsymbol\ levels of \ba\ were measured with the same atomic ion and are shown in Fig. \ref{fig:2}. These spectra were obtained by using a modular digital synthesis platform \cite{hong:2015} to rapidly alternate between Doppler cooling and weak optical excitation for fluorescence spectroscopy to prevent laser-induced lineshape distortions \cite{wolf:2008}. All measurements have a $\pm$20 MHz systematic uncertainty primarily due to drift of the wavemeter used to stabilize the lasers. To measure the \psymbol\ hyperfine splitting (Fig. \ref{fig:1}a), a laser sideband frequency near the \psymbol, $F$ = 1 $\leftrightarrow$ \ssymbol, $F$ = 1 transition is scanned. When this frequency is near resonance, and without laser frequency $\nu_{1}^{r}$, the population of the \dsymbol, $F$ = 2 states is increased. We utilize the resulting decrease in fluorescence to measure the \psymbol\ hyperfine splitting $\Delta_2 = 1840(2)_{\text{stat}}$ MHz (see Fig. \ref{fig:2}b). To measure the \ssymbol\ hyperfine qubit splitting,  the laser sideband $\nu_{1}^{b}$ near the \psymbol, $F = 1$ $\leftrightarrow$ \ssymbol, $F=0$  transition is scanned. The fluorescence is maximized when $2\nu_0 = \Delta_1 + \Delta_2$ (see Fig. \ref{fig:2}a). We measure the hyperfine qubit splitting $\Delta_1 = 9931(2)_{\text{stat}}$ MHz. In order to measure the \dsymbol\ hyperfine splitting, we increase the population of the \dsymbol, $F = 2$ manifold by applying a laser sideband at frequency $\nu_{0}^b - \Delta_2$. The fluorescence is maximized when the laser sideband $\nu_{1}^r = \nu_{0}^{r}+\Delta_3 - \Delta_2$ (see Fig. \ref{fig:2}c). We measure $\Delta_3 = 937(3)_{\text{stat}}$ MHz.

Efficient laser cooling of the ion also requires knowledge of the electronic transition frequencies. We measure these transitions in \ba\ using the values of the measured hyperfine splitting and scanning $\nu_{0}^{b}$ and $\nu_{0}^{r}$. Defining the isotope shift of the $\textit{i}$-th electronic transition $\delta \nu_{A,138}^{i} \equiv \nu_{A}^i - \nu_{138}^i$, with $\textit{i}$ = $b$ ($r$) for the transitions near 493~nm (650~nm), we measure the isotope shifts in \ba\  to obtain $\delta \nu^{b} = 355(4)_{\text{stat}}$ MHz  and $\delta \nu^{r} =  198(4)_{\text{stat}}$ MHz.

With these data, the transition frequencies necessary for laser cooling and hyperfine qubit operation are now characterized for all isotopes of $\text{Ba}^+$ with half-life greater than a few weeks, and are shown in Table \ref{tab:iso shifts}. Since all of these transitions are resolved and are simultaneously addressable using a broadband, fiber-coupled EOM,  isotopic purification is possible in situ through laser heating. This allows for the production of single-species Coulomb crystals, even for trace species, as shown in Fig. \ref{fig:1}b.

Finally, the isotope shifts can be decomposed into two terms
\begin{equation}\label{eq:iso_shift}
\delta \nu_{A,A'}^i = k_{\text{MS}}^i \Big (\frac{1}{A}-\frac{1}{{A'}} \Big ) + F_i \lambda_{A,A'}
\end{equation}
where $k_{\text{MS}}$ is the sum of the normal and specific mass shifts, $F_i$ is the field shift \cite{heilig:1974}, and $\lambda_{A,A'}$ is the Seltzer moment of the nuclei of isotopes $A$ and $A'$  \cite{blundell:1987}. To lowest order, the Seltzer moment $\lambda_{A,A'}$ is equal to the difference in the mean of the squared nuclear charge radii of an isotope pair: $\delta \langle r^2 \rangle_{A,A'}=\langle r_{A}^2 \rangle - \langle r_{A'}^2 \rangle$  \cite{heilig:1974, blundell:1987}. 

\begin{figure}[t!]
\includegraphics[width=0.95\columnwidth]{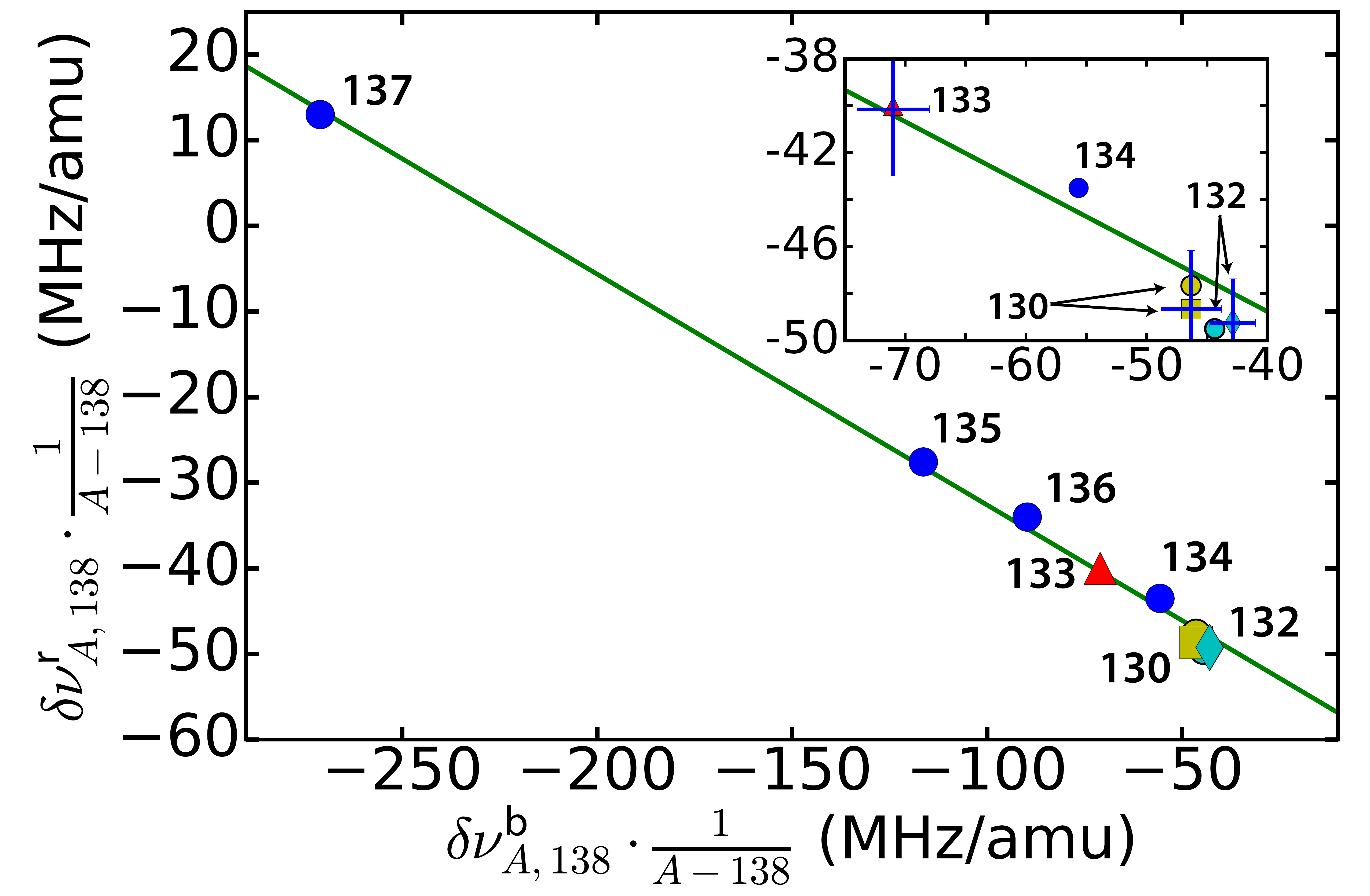}
\caption{\label{fig:3} King plot for the \psymbol\ $\leftrightarrow$ \dsymbol\  electronic transition as a function of the \ssymbol\ $\leftrightarrow$ \psymbol\  electronic transition. Each point represents a pair of barium isotopes ($^A$Ba$^+$,$^{138}$Ba$^+$), labeled by $A$, where the frequency shift is normalized by the mass difference $(\nu_A-\nu_{138})/(A-138)$. Red triangle: $^{133}$Ba$^+$, yellow square:  $^{132}$Ba$^+$, and cyan diamond:  $^{130}$Ba$^+$ include spectroscopic measurements from this work. Blue circles are reported isotope shifts taken from references \cite{wendt:1984, villemoes:1993}. Yellow and cyan circles are derived from reported isotope shifts of the \ssymbol\ $\leftrightarrow$ \psymbol\ transition \citep{wendt:1984}, and calculated isotope shifts of the \psymbol\ $\leftrightarrow$ \dsymbol\ transition \cite{wendt:1984, vanhove:1982}.}
\end{figure}

Following  Eqn. \ref{eq:iso_shift}, a King plot, shown in Fig. \ref{fig:3},  summarizes spectroscopic data for barium atomic ions  along with our measurements of $\delta \nu_{130,138}^{r}$, $\delta \nu_{132,138}^{r}$, and $\delta \nu_{133,138}^{r}$. Using previous spectroscopic data \cite{hohle:1976, wendt:1984, villemoes:1993}, the fitted slope of -0.26  is close to a theoretical calculation of the slope -0.288 \cite{olsson:1988}. The fitted slope, field and specific mass shifts of 988 MHz/$\text{fm}^2$ and 360~MHz respectively \cite{fricke:1983, villemoes:1993}, and the new measurement of $\delta \nu_{133,138}^{r}$ are combined to determine $\delta \langle r^2 \rangle_{133,138} = -0.104~\text{fm}^2$.

In summary, we have demonstrated trapping of \ba\ atomic ions produced via laser ablation of a microgram source.  By leveraging the frequency selectivity of laser heating and cooling, we isotopically purify the trap sample to achieve efficient laser cooling of trapped \ba\ ions. Using the same, single trapped \ba\ ion we have measured the previously unknown \dsymbol\ hyperfine splitting $\Delta_3=937(3)_{\text{stat}}$  MHz and isotope shift of the \psymbol\ $\leftrightarrow$ \dsymbol\ transition $\delta \nu_{133,138}^{r} = 198 (4)_{\text{stat}}$   MHz. These measurements all have a $\pm$20 MHz systematic uncertainty. The determination of these spectroscopic values along with the methods we have presented for trap loading from micrograms of radioactive material should enable the use of \ba\ for trapped ion QIP.

The advantages that \ba\ ion qubits promise over other species used for QIP are largely due to a unique combination of a nearly ideal atomic structure and the wavelength constraints of practical optical systems. First, unlike other ions hosting $M = 0$ clock-state qubits, the optical transitions that must be addressed to use \ba\ are all in the visible and near IR, allowing the integration of photonic technologies -- such as the fiber EOMs used in this work and very long optical fibers for quantum communication -- that do not perform well in the UV. Second, the spin-1/2 nucleus of \ba\ produces a hyperfine clock-state qubit that can be initialized and detected quickly using frequency-selective optical transitions \cite{lee:2003, olmschenk:2007}. Third, the unusually long-lived \otherdsymbol\ state in \ba\ should allow both ultra-high fidelity state-selective shelving detection \cite{harty:2014} and clock-state optical qubit operation. Therefore, \ba\ ions provide robust hyperfine and optical frequency qubits in the same system, allowing use of the full suite of trapped ion entangling gates in a single species, and represent an attractive candidate for future trapped ion QIP. 

This work was supported by the US Army Research Office under award W911NF-15-1-0273. We thank Rainer Blatt, Jungsang Kim, Michael Mills, Chris Monroe, and Prateek Puri for helpful discussions. We thank Tyler Jackson, Saed Mirzadeh, Anthony Ransford, Christian Schneider, Calvin Ye, and Peter Yu for technical assistance.

\bibliography{Ba133bib}

\end{document}